# Chiral electroluminescence from thin-film perovskite metacavities

Seongheon Kim[1], Soo-Chan An[1], Younggon Kim[1], Yun Seop Shin[2], Alexander A. Antonov[3], In Cheol Seo[1], Byung Hoon Woo[1], Yeonsoo Lim[1], Maxim V. Gorkunov[3,4,*], Yuri S. Kivshar[5,*], Jin Young Kim[2,*], Young Chul Jun[1,6,*]

[1]Department of Materials Science and Engineering, Ulsan National Institute of Science and Technology (UNIST), Ulsan 44919, Republic of Korea

[2]School of Energy and Chemical Engineering, UNIST, Ulsan 44919, Republic of Korea

[3]Shubnikov Institute of Crystallography of the Federal Scientific Research Centre "Crystallography and Photonics", Russian Academy of Science, Moscow 119333, Russia

[4]National Research Nuclear University MEPhI (Moscow Engineering Physics Institute), 115409 Moscow, Russia,

[5]Nonlinear Physics Centre, Research School of Physics, Australian National University, Canberra ACT 2601, Australia

[6]Graduate School of Semiconductor Materials and Devices Engineering, UNIST, Ulsan 44919, Republic of Korea

(Dated: February 10, 2023)

*Email: (YCJ) ycjun@unist.ac.kr, (JYK) jykim@unist.ac.kr, (MVG) gorkunov@crys.ras.ru, (YSK) yuri.kivshar@anu.edu.au

**ABSTRACT**

Chiral light sources realized in ultracompact device platforms are highly desirable for various applications. Among active media employed for thin-film emission devices, lead-halide perovskites have been extensively studied for photoluminescence due to their exceptional properties. However, up to date, there have been no demonstrations of chiral electroluminescence with a substantial degree of circular polarization (DCP), being critical for the development of practical devices. Here, we propose a new concept of chiral light sources based on a thin-film perovskite metacavity and experimentally demonstrate chiral electroluminescence with DCP approaching 0.38. We design a metacavity created by a metal and a dielectric metasurface supporting photonic eigenstates with close-to-maximum chiral response. Chiral cavity modes facilitate asymmetric electroluminescence of pairs of left and right circularly polarized waves propagating in the opposite oblique directions. The proposed ultracompact light sources are especially advantageous for many applications requiring chiral light beams of both helicities.

**INTRODUCTION**

Chirality – non-equivalence to mirror images – is optically manifested by different response to left circularly polarized (LCP) and right circularly polarized (RCP) light (*1–3*). The basic chiral phenomena exhibited by natural materials are circular dichroism and polarization rotation which are both typically very weak. To overcome these limitations, various nanostructures, including three-dimensional chiral structures (e.g., plasmonic helices) and two-dimensional planar chiral structures (e.g., gammadions) have been extensively studied (*4–7*). Strongly enhanced chiral light–matter interactions in these nanostructures can be utilized for important applications (*8–13*). Particularly, ultracompact chiral light sources with a large degree of circular polarization (DCP) are highly desirable for displays, optical recording, optical communication, bioimaging, and biomedical diagnosis (*14–19*).

Thin-film light-emitting diodes (LEDs) have emerged as a key technology for compact light sources (*20–22*). Among thin-film LED materials, lead halide perovskites have demonstrated remarkable properties, such as high photoluminescence (PL) quantum yields, narrow emission width (or high color purity), and high charge-carrier mobility (*23–26*). In addition, widely tunable bandgap energy makes perovskite materials ideal for light sources in the visible spectral range. Recently, chiral light sources based on perovskite materials have attracted considerable attention to fully exploit their exceptional material properties. Extensive research has been conducted to obtain chiral PL from perovskite materials (*27–35*). For example, chiral perovskite materials can be created by incorporating chiral organic ligands on the surface of perovskite crystals or intercalating chiral organic cations into the perovskite lattice (*27–29*). However, achieving large DCP in such chiral perovskite materials at room temperature remains challenging. More importantly, chiral electroluminescence (EL) with a large DCP, critical for compact device implementations, has not been demonstrated yet.

Here, we propose, design, fabricate, and demonstrate experimentally a novel thin-film perovskite ($CsPbBr_3$) metacavity with DCP of chiral EL approaching 0.38 at room temperature, i.e., significantly larger than all previously reported cases (*36–38*). Our design employs a layer of *achiral* perovskite confined between metal and a metasurface created by an array of triangular silicon particles with broken inversion symmetry. Chiral Fabry–Pérot modes supported by the metacavity determine asymmetry of EL, and pairs of LCP and RCP waves of equal powers are generated in the opposite oblique directions. Our theoretical analysis and modeling reveal a crucial role of close-to-maximum chiral metamirror eigenstates underpinning the EL chirality. Our experimental results suggest employing such metacavities as ultracompact sources of chiral light.

**RESULTS**

**Concept of thin film metacavity**

The purpose of the metacavity is to enhance EL from an active perovskite emission layer and to produce high chiral asymmetry, i.e., to support emission of either LCP or RCP waves in oblique directions. As shown in Fig. 1A, the metacavity is created by stacking the perovskite emission layer between a metal mirror and a specifically designed metamirror. We omit the charge transport interlayers in Fig. 1A for clarity. Optical role of the metamirror is twofold: it performs as a semi-transparent mirror supporting Fabry–Pérot cavity modes emitting waves of one circular polarization; and it suppresses Fabry–Pérot cavity modes of the opposite handedness emitting light of the opposite polarization. The key precondition for such peculiar functionality is the metamirror ability to selectively interact with circularly polarized waves. Various recent dielectric metasurfaces built as arrays of particles of high-refractive-index materials have repeatedly demonstrated that such selectivity can be achieved if their eigenstates are adjusted to the so-called maximum chirality (*39, 40*). For the normal light incidence, this requires nontrivial non-planar designs with variable height. For obliquely incident waves, simpler planar structures with broken inversion symmetry can perform in the extrinsic-chiral regime (*41*).

For a particular realization of the chiral cavity, we choose a square lattice of triangular polycrystalline silicon (poly-Si) particles, see the inset in Fig. 1A. With the dimensions properly adjusted (L = 200 nm, $t_{Si}$ = 40 nm, P = 270 nm), they host photonic eigenstates with a narrowband spectrum crossing the $CsPbBr_3$ emission band (wavelengths ~520 nm) as shown in Fig. 1B. Apart from the close-to-normal directions, the eigenstate polarization is very close to ideal circular in a remarkably wide range of the angles of incidence. This close-to-maximum chiral eigenstate together with the background metasurface reflection contributes to the formation of the metacavity resonance in the extrinsic-chiral regime (see also Note 1). For an oblique direction, the metacavity can support a Fabry–Pérot resonance and corresponding field enhancement *inside* the cavity only for EL of one circular polarization. For the opposite oblique direction, the situation is reversed, and the EL of the opposite circular polarization is obtained. Note that, in the directions not far from the normal, both metal and meta- mirrors flip circular polarizations upon reflection (discussed in more detail below), so that the cavity modes of the opposite handedness remain uncoupled. In a more general context, the flipping makes the cavity qualitatively different from those built of pairs of helicity-preserving metamirrors (*42–44*).

The DCP of EL is defined by the intensities of emitted circularly polarized waves as $\rho_c = (I_{RCP} - I_{LCP})/(I_{RCP} + I_{LCP})$ and it can be conveniently evaluated according to the reciprocity principle. Indeed, the power and polarization in the far field of an ensemble of randomly positioned and oriented dipole sources can be obtained by solving the reciprocal problem for the *incidence* of correspondingly polarized and directed plane waves and by evaluating the field intensity enhancement factors $I^{inc}_{RCP,LCP}/I_0$ as integrals over all dipole positions (see Methods). Accordingly, we simulate RCP and LCP waves incident onto the meta-cavity at different angles, integrate the intensity of the electric fields in the perovskite emitting layer and obtain the enhancement factor for each angle. The colormaps in Fig. 1C show the LCP and RCP intensity enhancement factors (cavity thickness $d = 150$ nm). The colormaps are given in the axes of free-space light wavelength $\lambda$ and the in-plane wavevector component $k_x/k_0$, where $k_0 = 2\pi/\lambda$. The corresponding DCP colormap is also shown in Fig. 1D. The shaded regions on all colormaps represent the CsPbBr$_3$ emission band near a 520 nm wavelength and they nicely cover the range of the strongest EL chirality (maximum DCP). Therefore, efficient implementations of the meta-cavity concept require fine spectral adjustment of the metasurface eigenstates, cavity modes, and the emission band. In the following, we account the theory behind the main design steps.

**Coupled-mode theory (CMT) of chiral metacavity**

CMT phenomenology describes resonant response of metasurfaces including those with intrinsic optical chirality (*39, 40, 44, 45*). It can be generalized to account for the extrinsic chiral effects as light is obliquely incident upon a metamirror with broken inversion symmetry (see Note 2 for details). The corresponding co-polarized reflection and cross-polarized transmission coefficients read as:

$$r_{RR} = r_{LL} = t_{RL} = t_{LR} = -\frac{m_R m_L}{i(\omega-\omega_0)-\gamma_0}, \quad (1)$$

while the cross-polarized reflection and co-polarized transmission coefficients are:

$$r_{RL} = \rho - \frac{m_L^2}{i(\omega-\omega_0)-\gamma_0}, \quad r_{LR} = \rho - \frac{m_R^2}{i(\omega-\omega_0)-\gamma_0}, \quad (2)$$

$$t_{RR} = \tau - \frac{m_R^2}{i(\omega-\omega_0)-\gamma_0}, \quad t_{LL} = \tau - \frac{m_L^2}{i(\omega-\omega_0)-\gamma_0}, \quad (3)$$

where $r_{ij}$ and $t_{ij}$ indicate the reflection and transmission coefficients of the output polarization $i$ from the input polarization $j$, respectively, and $\tau$ and $\rho$ are background transmission and reflection coefficients. The resonant terms correspond to the excitation of a photonic eigenstate, and $\omega_0$ is its frequency, $\gamma_0$ is its decay rate, and $m_{R,L}$ are the parameters of the eigenstate coupling to RCP and LCP waves, respectively. For a known distribution of the eigenstate displacement current density $\mathbf{J}(\mathbf{r})$, the coupling parameters are proportional to the overlap integrals (*39*):

$$m_{R,L} \propto \int_V \mathbf{J}(\mathbf{r}) \cdot \mathbf{e}_{R,L}\, e^{i\mathbf{k}\cdot\mathbf{r}}\, dV, \tag{4}$$

where $\mathbf{e}_{R,L}$ is the electric field unit vector of an RCP or LCP wave and $\mathbf{k}$ is its wavevector.

To decouple the state from RCP waves, it is sufficient to set $m_R = 0$ which establishes the *maximum extrinsic chirality*. As seen from the diagram in Fig. 1B, the meta-mirror eigenstate indeed closely approaches this condition in a wide range of the angles of incidence. To quantitatively analyze the underlying mechanism, we apply a multipole expansion to Eq. (4) and express it as (Note 3):

$$m_{R,L} \propto P_x \cos\theta_s \mp iP_y \pm i\frac{k_x}{\omega} M_z - \frac{k_x}{6}\left[\cos\theta_s\,(Q_{zz} - Q_{xx}) \pm iQ_{xy}\right], \tag{5}$$

where $P_i$, $M_i$, and $Q_{ij}$ are the Cartesian components of the eigenstate electric dipole, magnetic dipole, and electric quadrupole moments, respectively, and $\theta_s$ is the angle of incidence from the glass substrate. Numerically obtaining the eigenstate current distribution $\mathbf{J}(\mathbf{r})$ we evaluate and compare the relative magnitudes and phases of the terms entering Eq. (5) in Figs. 2A and 2B, correspondingly. One can see that the diagonal quadrupole components contribute negligibly, $iP_y$ and $iM_z$ contribute almost exactly in-phase with $P_x$, while $iQ_{xy}$ contributes with a relative $\pi/2$ phase shift. Therefore, the condition $m_R = 0$ can be approached as a result of the mutual compensation of the first three dipolar terms in Eq. (5).

To quantify the degree of eigenstate coupling selectivity, we introduce the eigenstate circular dichroism defined as:

$$CD_m = \frac{|m_L|^2 - |m_R|^2}{|m_L|^2 + |m_R|^2}. \tag{6}$$

Comparing its values obtained using Eqs. (4) and (5), as well as using the reduced multipole expansion accounting only for the main contributions (Note 3), we find that the eigenstate indeed closely approaches the maximum chirality, as its CD stays close to unity in a very broad range of the angles of incidence (see Fig. 2C). Intriguingly, the multipolar content of the state varies with the angle. For smaller angles, the magnetic dipole contribution is almost negligible, and the chirality is provided by the interplay of the electric dipole components $P_x$ and $P_y$ which are shifted by $\pi/2$ in phase. At larger angles, the magnetic dipole moment contributes more substantially, although both components of the electric dipole also play similarly important roles.

Knowing the metamirror characteristics, one can obtain analytical expressions for the fields inside the metacavity and evaluate the intensity enhancement factors. As both the metamirror and the metal mirror reflect LCP/RCP light flipping its helicity and the metamirror transmits preserving the helicity, the cavity modes of opposite handedness remain uncoupled. The LCP and RCP enhancement factors are obtained from two independent problems (Note 2) as proportional to:

$$\frac{I_{\text{LCP}}^{\text{inc}}}{I_0} \propto \frac{|t_{LL}|^2}{|1-r_{RL}r_{\text{metal}}\,e^{2ik_zd}|^2}, \quad \frac{I_{\text{RCP}}^{\text{inc}}}{I_0} \propto \frac{|t_{RR}|^2}{|1-r_{LR}r_{\text{metal}}\,e^{2ik_zd}|^2}, \tag{7}$$

where $r_{\text{metal}}$ is the reflection coefficient of the metal mirror and $k_z$ is the vertical wavevector component. One can readily employ Eq. (7) to explain the chiral asymmetry of field enhancement and, correspondingly, high DCP of EL. For example, for the exemplary wavelength and direction indicated by the black star in Figs. 1B–D, the transmission and reflection coefficients are given in fig. S8, where $|t_{LL}|$ has a resonance dip around 520 nm (therefore, $|t_{RR}| > |t_{LL}|$). In addition, the phases of $r_{RL}$ and $r_{LR}$ are significantly different, and thus the denominator in Eq. (7) also contributes to the field intensity and DCP. In general, depending on the cavity thickness $d$, a very different type of interplay between the resonant reflection (determined by $m_L$ and $m_R$) and the broad background reflection (determined by $\rho$) from the metamirror appears (see figs. S2C and S3). This metacavity response can be utilized to optimize the emission intensity and DCP of chiral EL.

**Sample fabrication and reflectance measurements**
In our EL device, a square lattice of poly-Si patterns is integrated with a perovskite EL device. Perovskite EL was measured on the transparent substrate side (fig. S5a). First, arrays of poly-Si patterns were created on a fused quartz substrate using electron-beam lithography. Each array has a different pattern shape (triangle or circle) and lattice period. figure S11 shows the optical microscopy and scanning electron microscopy images of the poly-Si patterns. On top of the poly-Si patterns, a 60-nm-thick indium tin oxide (ITO) layer was first deposited. Then, a hole transport layer and inorganic polycrystalline perovskite emitter layer ($CsPbBr_3$) were spin-coated in succession (*46, 47*). Finally, an electron transport layer and metal contact (LiF/Al) were sequentially deposited using a thermal evaporator. The measured cavity thickness was $d \sim 150$ nm. The top metal contact has a circular shape with 2 mm diameter and includes arrays of poly-Si patterns underneath. Figure S12A shows the PL and EL spectra from a reference sample without poly-Si patterns; $CsPbBr_3$ has an emission peak near 520 nm with a full width at half-maximum (FWHM) of 18 nm. The narrow emission width of perovskite materials can be advantageous because of the limited spectral bandwidth in chiral responses. Figure S12B shows the absorbance spectrum of a bare $CsPbBr_3$ film spin-coated on a quartz substrate, and fig. S12C presents the refractive index of the poly-Si film measured using ellipsometry.

To experimentally characterize the chiral response of our device, we first measured the angle-resolved reflectance spectrum using a custom-built Fourier-plane (or momentum-space) measurement setup (fig. S14A). CP light was incident from the substrate side, and the total reflection intensity was measured (fig. S15). A clear chiral response appears in the cavity with the triangular pattern; the LCP and RCP reflectance maps show distinctly different mode dispersions near the emission band (~520 nm). Using momentum-space tomography, we also constructed the cavity mode dispersion in the entire energy-momentum space (within the numerical aperture of a microscope objective) and further confirmed the chiral response of the metacavity resonances (fig. S16).

**Measurements of EL and PL**

To characterize the emission properties of our device, we measured EL and PL spectra in the Fourier plane. The chiral responses of the thin-film cavities with the triangular and circular patterns were compared. The LCP and RCP emission spectra were measured using a quarter-waveplate and linear polarizer. Because the circular metal contact (diameter: 2 mm) includes arrays of the poly-Si patterns underneath, a pinhole was placed in the real-space plane for the EL measurements to collect emission from the patterned region only (60 μm × 60 μm in size for each array) (fig. S14B). The emission spectrum was then measured in the Fourier plane.

We directly compare EL and PL spectra to characterize the emission properties of our device. Figure 3 presents the EL and PL DCP maps as a function of $\lambda$ and $k_x/k_0$. Because the DCP is the ratio of the RCP and LCP emission intensities, the DCP map becomes noisy outside of the emission band owing to small emission intensities. The upper row in Fig. 3A shows the EL DCP maps from the triangular pattern for the periods of 250, 260, and 270 nm, respectively (fig. S18 shows the separate RCP and LCP emission spectra for the 270 nm period). EL from the triangular pattern shows clear chiral responses, which agrees with our reciprocity calculations in Fig. 1d. The chiral emission band gradually redshifts with the increase in the pattern period. The PL spectrum (the lower row in Fig. 3A) also shows the same behavior. Figures 3C and 3D show the EL line spectra and the corresponding DCP at $k_x/k_0 = -0.27$ and $k_x/k_0 = +0.27$ (period 270 nm), respectively. The EL spectrum shows a clear chiral response; the LCP EL intensity is larger for $k_x < 0$ while the RCP EL intensity is larger for $k_x > 0$. The peak DCP reaches $\rho_c \approx 0.38$, which is the largest EL DCP in perovskite materials to our knowledge. In contrast, EL and PL from the circular pattern (diameter: 150 nm) do not show such a chiral response along the $k_x$ direction (Fig. 3B).

We also conducted emission measurements over the $(k_x, k_y)$ plane (or Fourier plane) and obtained EL and PL images. Each point in the $(k_x, k_y)$ plane corresponds to a different far-field emission angle. Therefore, the EL and PL images in the Fourier plane indicate the far-field emission pattern from our device. Figure 4A presents a comparison of the LCP and RCP components for the EL (upper row) and PL (lower row) from the thin-film metacavity with the triangular pattern. The EL and PL images were measured using a 10 nm bandpass filter at 520 nm. In both EL and PL images, the LCP and RCP emissions are stronger for the negative $k_x$ and positive $k_x$ regions, respectively. This means that the LCP and RCP emissions are emitted into opposite angles. The angular plot of EL along the $k_x$ direction ($k_y = 0$) confirms this behavior (Fig. 4B). Figure 4A also shows the corresponding DCP of the EL and PL. The DCP $\rho_c$ in the ($k_x, k_y$) plane exhibits clear chiral responses in both cases. All these emission measurements agree with the cavity mode properties determined from the momentum-space tomography (reflectance) measurements (fig. S16). The angular plot of the LCP and RCP emissions obtained from the reciprocity calculation is also provided in Fig. 4B; we find that the experiment (upper panel) and reciprocity calculation (lower panel) are in good agreement overall.

We also note that the dark emission line (or suppressed emission) in EL and PL images appears on the opposite side of the bright (or enhanced) emission along the $k_x$ direction (Fig. 4B), which increases the contrast between the LCP and RCP emissions. Because DCP is the ratio of the LCP and RCP emission intensities, DCP in our device is further enhanced by this suppressed emission. The observed suppressed emission agrees well with the reciprocity calculations in Fig. 1C. Our experimental and theoretical studies demonstrate that Fabry–Pérot cavities with tailored mirror responses can be exploited in a clever way to improve the characteristics of the chiral light sources.

**DISCUSSION**

Our approach is generally applicable to other thin-film LEDs with different emissive materials such as quantum dots (*20*). In our concept and its realization, chiral EL with a substantial value of DCP is achieved from *achiral emitters* without using spin injection or filters (*36, 38*). We have experimentally confirmed that both EL and PL spectra show the same types of the chiral response being determined by the chiral behavior of the metacavity resonance.

We have clarified theoretically that nearly maximum chirality of the inversion-symmetry-broken metamirror results from the interplay of the *three dipole moments*: two electric dipoles ($P_x$ and $P_y$) and one magnetic dipole ($M_z$). In addition, we have explained the physical mechanism of the chirality asymmetry and the field enhancement in thin film metacavities, thus providing the general strategies for a design of compact EL devices that can enhance simultaneously chiral emission intensity and DCP.

Thus, we have proposed and demonstrated experimentally *a new type of light source* that exhibits chiral EL in a thin-film device platform. Many chiral applications require both LCP and RCP light of the same power (e.g., circular dichroism and enantiomer selection), and thus our studies provide novel opportunities for innovative device integration and applications requiring ultracompact sources of chiral light.

## MATERIALS AND METHODS

### Materials and perovskite synthesis

Cesium bromide (CsBr, 99.999%) and dimethyl sulfoxide (DMSO, 99.9%) were purchased from Sigma-Aldrich. Lead bromide ($PbBr_2$, for perovskite precursors) was purchased from Tokyo Chemical Industry Co. Ltd., and 1,3,5-tris(*N*-phenylbenzimidazol-2-yl) benzene (TPBi, 99.9%) was purchased from OSM. All the chemical materials were used as received without further purification. The precursor solution (0.2 M) for the $CsPbBr_3$ perovskite film was prepared by dissolving CsBr and $PbBr_2$ (molar ratio, 1.7:1.0) in anhydrous DMSO under continuous stirring overnight at 60 °C. The perovskite precursor solution was filtered with a pore size of 0.20 µm before use. The polycrystalline perovskite films were obtained by spin-coating at 4000 rpm for 60 s and the resulting films were annealed at 100 °C for 10 min.

### Fabrication of poly-Si patterns

A polycrystalline silicon (poly-Si) film (thickness: 40 nm) was deposited on a fused (amorphous) quartz substrate using low-pressure chemical vapor deposition (LPCVD) at 565 °C, followed by annealing for 1 hr at 950 °C. Arrays of triangular and circular patterns were created in the poly-Si film using electron-beam lithography and dry etching. The side length *L* of the triangular pattern is 200 nm while the diameter of the circular pattern is 150 nm. The lattice period is varied from 250 nm to 270 nm. The patterned arrays were first created on a PMMA resist (AR-P 6200.09). The size of each array was 60 µm × 60 µm. After developing the patterned resist, a Cr thin film was deposited as a hard mask for dry etching using an electron beam evaporator, followed by a lift-off process. The inductively coupled plasma reactive ion etching was then conducted using an Ar and $Cl_2$ gas mixture as the main etchant to create patterns in the poly-Si film. Finally, the Cr hard mask was removed using a Cr etchant purchased from Sigma-Aldrich.

### EL device fabrication

To fabricate an EL device, a 60-nm-thick indium tin oxide (ITO) film was first deposited on the patterned poly-Si structure. The sample was then sequentially cleaned by ultrasonication in deionized water, acetone, and isopropyl alcohol and dried in an oven overnight at 80 °C. The substrate was then treated with UV ozone for 30 min before spin-coating of poly(3,4-ethylenedioxythiophene):poly(styrene sulfonate) (PEDOT:PSS). The PEDOT:PSS solution was spin-coated at 3000 rpm for 40 s and then annealed at 140 °C for 15 min. After spin-coating of the $CsPbBr_3$ film, TPBi (30 nm), LiF (1 nm), and Al (100 nm) were sequentially deposited by thermal evaporation under high vacuum ($< 10^{-6}$ Torr). The thickness and structural profile of each layer were determined using atomic force microscopy (fig. S13).

**Reflectance measurement and momentum-space tomography**

The dispersion of the cavity modes in our device was measured in a custom-built Fourier-plane setup. To conduct angle-resolved reflectance measurements, the back focal plane of the microscope objective was imaged instead of the device sample surface. The spectrum in the Fourier plane was measured using a monochromator slit as the line aperture in the Fourier plane (see the setup schematic in fig. S14). Both reflectance and luminescence measurements were performed from the quartz side. The in-plane wavevector $k_x$ is defined as $k_x = k_0 \sin\theta$, where $\theta$ is the incidence angle along the x-axis. Note that refraction occurs at the quartz/air interface; thus, the internal and external angles are different in our case. Throughout this study, we used the external angles measured in air.

The chiral responses of the cavity modes were characterized using reflectance measurements. Collimated white light was circularly polarized using a broadband Glan Thompson linear polarizer and quarter waveplate and focused on the sample surface using a high numerical aperture (NA) microscope objective (NA: 0.75). The total intensity of the reflected beam was then measured. The measured raw reflectance spectrum was normalized to the background reflectance of a flat silver film. For momentum-space tomography, the lens in front of the monochromator slit was gradually moved using a motorized microstage to construct the reflectance map in energy-momentum space.

**Measurements of luminescence spectra**

Angle-resolved luminescence spectrum measurements were also conducted using the Fourier-plane setup. The PL spectrum of the device was measured using a *linearly* polarized diode laser at 450 nm as an excitation light source. The circular polarization of PL and EL was measured using a linear polarizer and quarter waveplate. All measurements were conducted at room temperature. For the EL measurements, the bottom ITO film was used as an anode and biased against the circular top metal contact. A continuous bias (4 ~ 4.5 V) was applied to measure EL. The EL from each pattern array (60 μm × 60 μm in size) was extracted using real-space filtering for spectral measurements. The emission DCP $\rho_c$ is defined as the ratio between the RCP and LCP emission intensities ($I_{RCP}$ and $I_{LCP}$, respectively): $\rho_c = (I_{RCP} - I_{LCP})/(I_{RCP} + I_{LCP})$. It is also related to the normalized Stokes parameter: $\rho_c = S_3/S_0$.

To obtain the Fourier-plane images of EL and PL (Fig. 4), the monochromator slit was fully opened, and the monochromator grating was set to the zero position. The Fourier plane was directly imaged on a two-dimensional charge-coupled device (CCD) camera. The far-field emission profile of the device can be measured in the Fourier ($k_x$-$k_y$) plane. The radius of the Fourier-plane image corresponds to the NA of the microscope objective. The EL and PL images were measured using a 10 nm bandpass filter at 520 nm. The bandpass filter covered the emission band near the peak emission wavelength.

**Reciprocity calculation of circularly polarized emission**

The reciprocity principle states that the source and detector of electromagnetic fields can be interchanged without affecting the physical situation (*48*). Therefore, the far-field radiation power and polarization from an ensemble of randomly positioned and oriented dipolar emitters can be obtained by simply calculating the field enhancements at the dipole positions under incident light (*49, 50*). The reciprocity principle ensures that the field enhancement calculation in the emission layer produces the same result as the dipole simulation. However, compared to typical dipole simulations, the reciprocity calculation simplifies emission simulations and also enables a deeper understanding of the emission process in our device because the field enhancement in a photonic cavity is more straightforward to understand and optimize.

The reciprocity principle states that, for any two current densities $\mathbf{j}_{1,2}$,

$$\int \mathbf{j}_1 \cdot \mathbf{E}_2 dV = \int \mathbf{j}_2 \cdot \mathbf{E}_1 dV \tag{8}$$

where $\mathbf{E}_{1,2}$ is the electric fields produced by $\mathbf{j}_{1,2}$. For a dipole source $\mathbf{p}_m$, where $\mathbf{j}_m = -i\omega \mathbf{p}_m \delta(\mathbf{r} - \mathbf{r}_m)$ ($m$ = 1, 2), this is simplified to

$$\mathbf{p}_1 \cdot \mathbf{E}_2(\mathbf{r}_1) = \mathbf{p}_2 \cdot \mathbf{E}_1(\mathbf{r}_2) \tag{9}$$

The DCP is then determined by the intensity enhancement factors $I^{\text{inc}}_{\text{RCP,LCP}}/I_0$ integrated across the perovskite emitting layer obtained by solving the reciprocal problem of illumination by reversely propagating incoming plane waves. For circular polarizations, the reciprocal problem with interchanged sources and detectors involves waves of the same helicity[45], but, additionally, interchanging obliquely incident and emitted waves implies reversing the in-plane wavevector component. In the considered geometry this is equivalent to the handedness inversion, and, for simplicity, we preserve the in-plane wavevector component and evaluate DCP as $\rho_c = \left(I^{\text{inc}}_{\text{LCP}} - I^{\text{inc}}_{\text{RCP}}\right)/\left(I^{\text{inc}}_{\text{LCP}} + I^{\text{inc}}_{\text{RCP}}\right)$.

**Metamirror eigenstate analysis**

The complex eigenfrequencies and polarization current distributions of the metasurface eigenstates are numerically obtained using COMSOL Multiphysics eigenvalue solver. The corresponding overlap integrals Eq. (4) as well the multipole moments entering Eq. (5) are numerically evaluated. To analyze the eigenstate polarization, we calculate the integrals $m_\alpha = \int J_\alpha(r) e^{i\mathbf{k}\cdot\mathbf{r}} dV$, $\alpha = x, y, z$, then extract the components of vector $\mathbf{m}$ orthogonal to $\mathbf{k}$ and consider them in the reference frame of obliquely incident wave as components of a Jones vector. The eigenstate coupling to RCP/LCP waves is described by $m_{R,L} = \mathbf{m} \cdot \mathbf{e}_{R,L}$, where $\mathbf{e}_{R,L} = (\mathbf{e}_{x'} \mp i\mathbf{e}_{y'})/\sqrt{2}$ is defined in the oblique Cartesian coordinate system $(x', y, z')$ with the $z'$-axis directed along $\mathbf{k}$. To visualize the eigenstate polarization in Fig. 1B, we use ellipses as symbols and their shape and inclination characterize the eigenstate polarization in a conventional way, while their size is inverse proportional to the eigenstate quality factor.

**Author contributions**

Y.C.J., J.Y.K, M.V.G., and Y.S.K. proposed the project. S.K., Y.K. and Y.S.S fabricated and characterized devices. S.K., I.C.S., B.H.W., and Y.L. conducted measurements. Y.C.J., S.K., and S.C.A. conducted analyses based on numerical simulations. M.V.G., A.A.A., and Y.S.K. conducted the CMT analysis and eigenmode simulation. All authors discussed the results and contributed to the manuscript preparation.

**Figures and Tables**

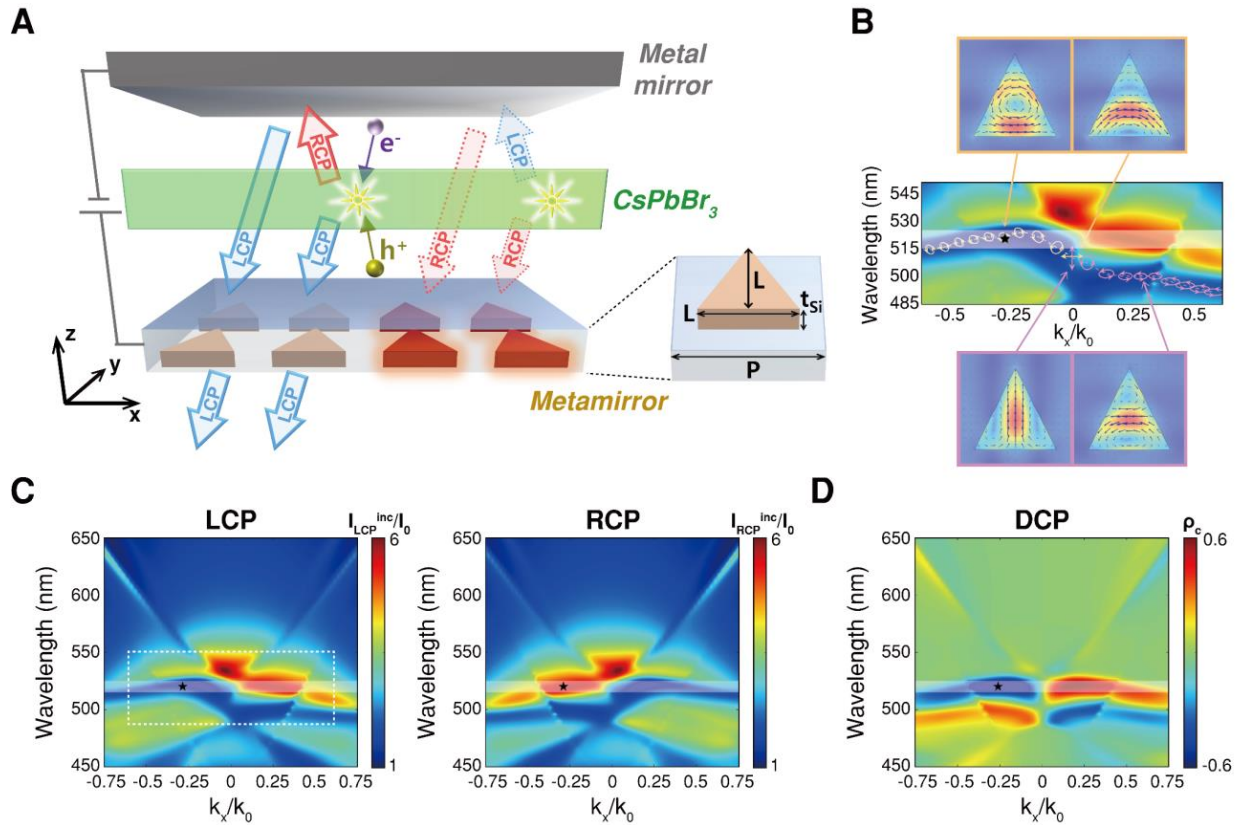

**Fig. 1. Concept of thin film metacavity for chiral electroluminescence.** (**A**) Schematic of the cavity formed by a metal mirror and a metamirror consisting of inversion-symmetry-broken poly-Si patterns immersed in a transparent ITO electrode. Inset: Square unit cell of the pattern containing a poly-Si triangle with the main dimensions indicated. For an oblique direction, the metacavity can support a Fabry–Pérot resonance for the light emission of one circular polarization while it suppresses the resonance for the other polarization. For the opposite oblique direction, the situation is reversed, and the emission of the opposite circular polarization is obtained. (**B**) Numerically obtained (see Methods) eigenstate spectra of the poly-Si triangular pattern with polarization ellipses shown on the line, superimposed on the LCP intensity enhancement colormap. Several representative polarization current distributions in the middle cut of a triangle for the linearly polarized (for normal incidence) and close-to-circularly polarized (for oblique incidence) eigenstates are given with the real and imaginary components of the current shown by red and blue arrows respectively. (**C**) Colormaps of intensity enhancement factor of LCP and RCP light (cavity thickness $d = 150$ nm) with the fragment shown in (B) indicated by the white dashed frame. (**D**) The corresponding DCP colormap. The simulated data are obtained for the realistic structure including all device layers (fig. S5). The emission band (515 – 525 nm) is indicated by shaded stripes in (B–D).

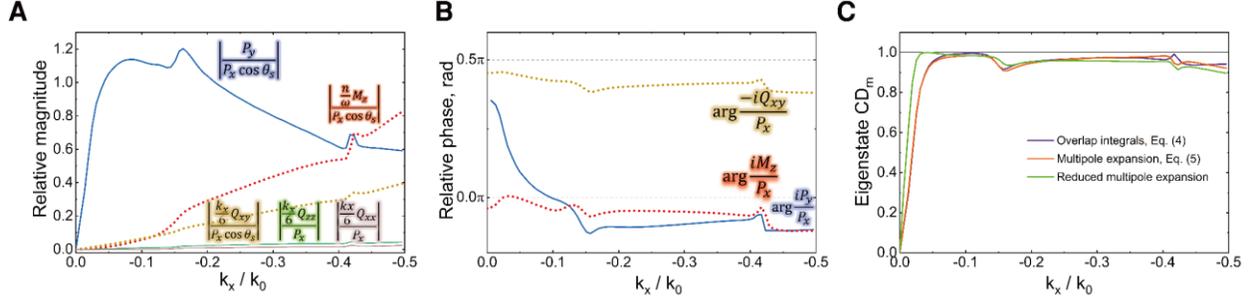

**Fig. 2. Multipolar mechanism of metamirror maximum chirality**. (**A**) Relative magnitudes of the multipolar terms contributing to the coupling parameters in Eq. (5). (**B**) Relative phases of the three substantial terms as counted from the phase of $P_x$. (**C**) Comparison of the eigenstate circular dichroism evaluated by the integrals Eq. (4), using the multipole expansion Eq. (5), and by the reduced expansion taking into account only the main contributions (Eq. (3.4) in Note 3).

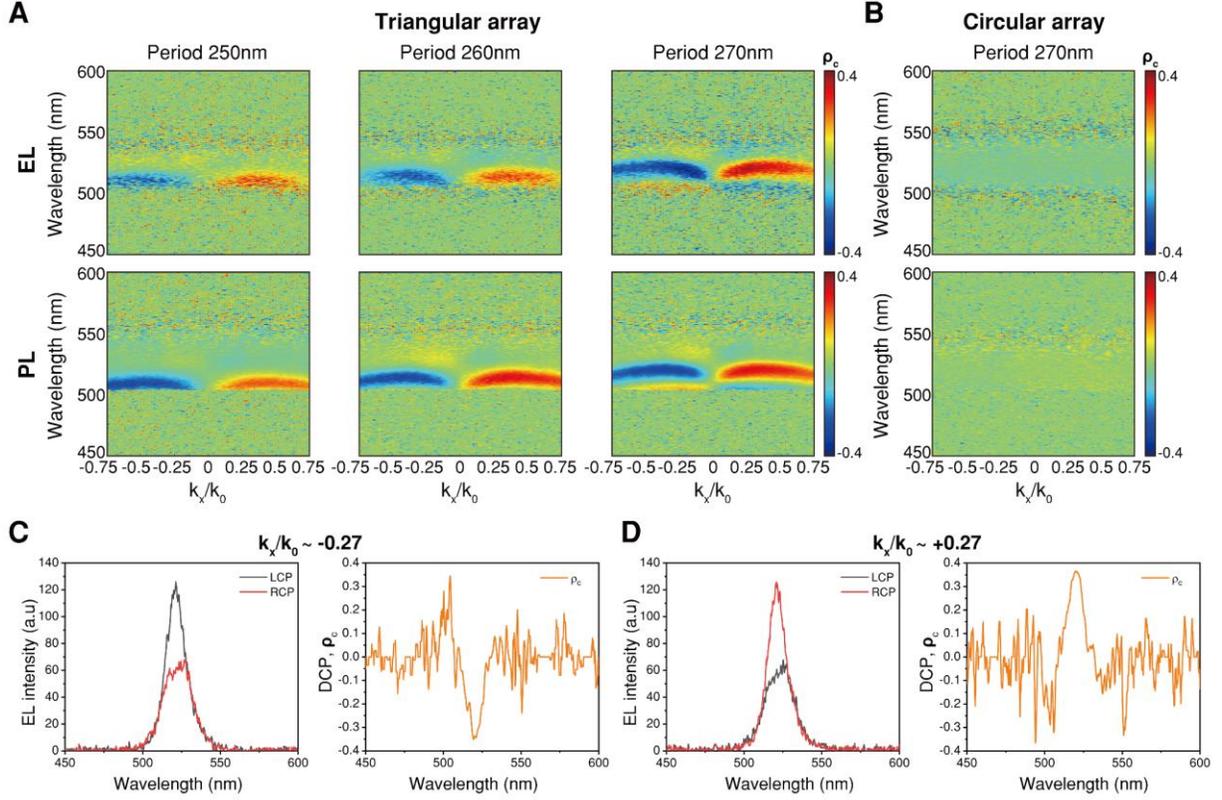

**Fig. 3. Experimentally measured EL and PL.** (**A**) EL and PL degree of circular polarization (DCP) for the triangular pattern of the different pattern periods (250 ~ 270 nm). The emission DCP is defined as the ratio between the RCP and LCP emission intensities ($I_{RCP}$ and $I_{LCP}$, respectively): $\rho_c = (I_{RCP} - I_{LCP})/(I_{RCP} + I_{LCP})$. Both EL and PL spectra show similar behavior. (**B**) EL and PL DCP for the circular pattern (period 270 nm). DCP is close to zero along the $k_x$ direction. (**C, D**) the EL line spectra and the corresponding DCP at $k_x/k_0 = -0.27$ (or $\theta = -15.7°$) and $k_x/k_0 = +0.27$ (or $\theta = +15.7°$), respectively. The peak EL DCP reaches ~0.38. The spectral width of the chiral response is wide enough to cover a large portion of the emission band.

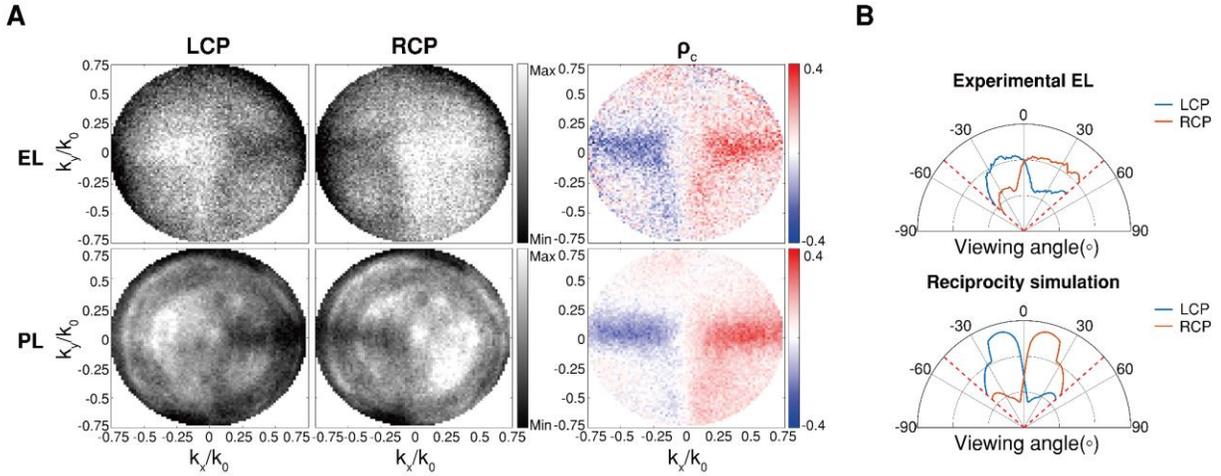

**Fig. 4. Fourier-plane images of EL and PL**. (**A**) The LCP and RCP emissions are measured in the $k_x$-$k_y$ plane using a 10 nm bandpass filter at 520 nm (near the peak emission wavelength). The EL and PL DCP ($\rho_c$) show a strong chiral response along the $k_x$ direction; LCP and RCP emissions are generated into opposite angles. (**B**) Angular distribution of the LCP and RCP emissions along the $k_x$ direction ($k_y = 0$). Top: experimentally measured EL. Bottom: reciprocity calculations (averaged in the wavelength range of 515 ~ 525 nm). The orange dotted lines indicate the detection limit due to the numerical aperture of a microscope objective (corresponding to 48.6°).